\documentstyle[prl,aps,epsfig]{revtex}
\begin{document}

\def\mev{\hbox{\ MeV}}
\def\lsim{\mathrel{\rlap{
\lower4pt\hbox{\hskip-3pt$\sim$}}
    \raise1pt\hbox{$<$}}}     
\def\gsim{\mathrel{\rlap{
\lower4pt\hbox{\hskip-3pt$\sim$}}
    \raise1pt\hbox{$>$}}}     

\draft


 \twocolumn[\hsize\textwidth\columnwidth\hsize  
 \csname @twocolumnfalse\endcsname              

\title{Fragile Signs of Criticality in the Nuclear Multifragmentation}

\author{K.K. Gudima\dag\ddag, M. P{\l}oszajczak\dag ~and
V.D. Toneev\dag\S }
\address{\dag\ Grand Acc\'{e}l\'{e}rateur National d'Ions Lourds,
CEA/DSM -- CNRS/IN2P3, BP 5027, F-14076 Caen Cedex 05, France}
\address{\ddag\ Institute of Applied Physics, Moldova Academy of
Sciences, MD-2028 Kishineu, Moldova}
\address{\S\  Bogoliubov Laboratory of Theoretical Physics,
Joint Institute for Nuclear Research, 141980 Dubna, Russia}

\date{\today}

\maketitle

\begin{abstract}
Deviations from an idealized equilibrium
phase transition picture in nuclear multifragmentation is studied in terms
of the entropic index. We investigate different heat-capacity features in the
canonical quantum statistical model of nuclear
multifragmentation generalized in the framework of Tsallis
nonextensive thermostatistics.
We find that the negative branch of heat capacity observed in quasi-peripheral
Au+Au collisions is caused primarily by the non-generic nonextensivity
effects.
\end{abstract}


 ]  

\narrowtext The nuclear multifragmentation process is studied in
the energetic collisions of heavy ions (HI). In these collisions,
strongly off-equilibrium transient system is formed which
equilibrates at the later stage of the reaction due to
dissipative processes. Perfectly equilibrated system, as assumed
in most theoretical descriptions of the multifragmentation decay
of the hot residue, is most probably never attained. This would
not be a serious problem if the nuclear fragmentation process
does not show any sign of the 'criticality' \cite{exp1,exp2}.
Indeed, the nonextensivity of weakly off-equilibrium finite
systems may qualitatively modify both the picture of the two phase
coexistence and signatures of the critical behavior in small
systems \cite{gppt00}. On the other hand, these nonextensivity
corrections to Boltzmann-Gibbs statistical mechanics (BGSM) have
no measurable effects on standard signatures of the equilibrium
such as the particle/fragment kinetic energy spectra or angular
distributions \cite{gppt00}. Neither the caloric curve nor the
negative heat-capacity branch measurements \cite{exp1,exp2}, both
put forward as an evidence for the nuclear liquid-gas phase
transition, can be interpreted unambiguously. The nonextensive
effects due to the long-range interaction/non-Markovian memory
effects or the multifractal phase boundary conditions
\cite{tsallis}, which  are crucial {\it only} in the 'critical
region' and cannot be reliably estimated, not only constitute an
integral part of the physics of HI collisions but also provide an
essential limitation to the understanding of the
multifragmentation based on the BGSM. The practical solution to
this problem would be to use those models of excited nuclear
matter which describe physics of limiting two phases outside of
the 'critical region' and to parameterize the critical region in
simple terms. It is the aim of this work to illustrate this
problematic in the thermodynamic (canonical) model of the
fragmentation which is extended to include the nonextensive
effects in the framework of the generalized thermostatistics
\cite{gppt00}.

A reasonable starting point could be any 'realistic' thermodynamic
fragmentation
model (for the list of examples see \cite{mult2,mult3}). This choice
offers several advantages, such as correct quantum statistics, correct
definition of fragment sizes and the fragment binding energies.
The Coulomb and surface effects can be tuned by analyzing
the observable quantities far outside of the 'critical region' and fragment
excitations can be included, if necessary.
Several models of this kind had an
unquestionable success in describing reaction products and their properties
from the regime of particle evaporation at low
excitation energies to the explosion at about 5 - 10 MeV/nucleon
\cite{mult2,mult3}. The new class of nonextensive thermodynamical models,
can be formulated in the framework of
the Tsallis generalized statistical mechanics (TGSM) \cite{tsallis}. TGSM
is based on an alternative definition for the
equilibrium entropy of a system whose $i$th microscopic state has probability
${\hat p}_i$ :
\begin{equation}
\label{eq1} S_q=k\frac{1-\sum_{i}{\hat p}_{i}^{q}}{q-1}
 =k\frac{\sum_{i}{\hat p}_{i}- \sum_{i}{\hat p}_{i}^{q}}{q-1} ~~,
\hspace{0.5cm} k>0
\end{equation}
where $q$ is the
entropic index and the normalization condition
\begin{equation}
\label{eq2}  \sum_{i}{\hat p}_i=1
\end{equation}
is used to get the second equality in (\ref{eq1}). The limit
$q=1$ corresponds to the BGSM. It is easy to verified that such
general properties as non-negativity, concavity and so on are
preserved by this new entropy definition. The main difference
between BGSM and TGSM is in the non-additivity of entropy in the
TGSM. Indeed, for two independent subsystems $A$, $B$, {\em i.e.}
such that the joint probability of $A+B$ is factorized into :
$\hat p_{A+B}=\hat p_A \hat p_B$, the global entropy in TGSM : $$
S_q(A+B)=S_q(A)+S_q(B)+(1-q) S_q(A) S_q(B)/k \ ,$$ is not equal
to the sum of the subsystem entropies.

The entropic index $q$ sometimes can be related in explicit way
to other basic quantities such as internal energy or free energy
of the system. Recently the Tsallis definition of entropy
(\ref{eq1}) was reinterpreted in terms of incomplete information
theory~\cite{Wang}. The condition (\ref{eq2}) means that all the
possible physical states are well-known and counted. But for
complicated systems we often in practice do not know all
interactions, cannot find the exact Hamiltonian, the exact
solution of equation of motion and the exact values of physical
quantities. Therefore, a part of information is lost and the
normalization (\ref{eq2}) is violated because the set of the
countable states becomes incomplete. By redefining the real
probability in Eq.(\ref{eq2}) to effective one  as $\hat p_i \to
\hat p_i^q$, one can keep the Tsallis definition of entropy (note
the 'invariance' of the numerator in Eq.(\ref{eq1}) with respect
to chosen normalization condition). Here the difference $q_1\equiv
q-1$ is related to the extra entropy of the system due to
neglected interactions. In this case the essential question is
not the preferential use of microcanonical ensemble (for which the
entropy is extremized with keeping only the normalization
restriction) or canonical one (when an additional restriction for
the energy conservation in average is applied) but rather the
point which interactions are included effectively by introducing
a free parameter $q$. The situation discussed we meet in the
multifragmentation where effects of thermal and chemical
non-equilibrium, slow expansion of decaying system  as well as
possible multifractal behavior of interphase boundary are out of
consideration in the BGSM.

Our further consideration is based on the canonical
multifragmentation model~\cite{pgt99}. The canonical ensemble
method in TGSM was introduced in \cite{tsallis98}. The main
ingredient of the nonextensive canonical quantum statistical
model of nuclear multifragmentation \cite{gppt00} is the
expression for the fragment partition function :
\begin{eqnarray}
\label{eq3}
{\omega}_q{(a,z)} &=& \sum_{\vec{p}}[1+q_1 \ \beta \
\varepsilon_{\vec{p}}{(a,z)} \ ]^{-1/q_1}
\end{eqnarray}
where $a$ and $z$ are the fragment mass number and the fragment
charge number, respectively. The fragment partition probability
equals :
\begin{eqnarray}
\label{eq4}
{\hat p}_{\vec {p}}{(a,z)}=[{\omega}_q{(a,z)}]^{-1}
[1+q_1 \ \beta \ \varepsilon_{\vec {p}}{(a,z)}]^{-1/q_1}
\end{eqnarray}
where $\varepsilon_{\vec {p}}{(a,z)}=p^2/2M+U{(a,z)}$ and $\beta
\equiv 1/T$. In the limit $q_1\rightarrow 0$, Eq.(\ref{eq4})
recovers the familiar expression : ${\hat p}_{\vec {p}}{(a,z)} =
 \exp{(-\beta \ \varepsilon_{\vec
{p}}{(a,z)}})/{\omega_1(a,z)}$. The internal energy $U$, which
includes the fragment binding energy and the fragment excitation
energy, the temperature-dependent surface energy,
 and the Coulomb interaction between fragments in the
Wigner-Seitz approximation, is parameterized as in \cite{mult3}.
 In the dilute gas approximation~\cite{bdg97}, the partition
function of a whole system   can be written as :
\begin{eqnarray}
\label{eq8}
{\cal Q}_{q}{(A,Z)} = \sum_{{\hat n} \in {\Pi}_{A,Z}}{\prod}_{a,z}
\frac{\left[ {\omega}_{q}{(a,z)}\right]^{N_{{\hat n}}{(a,z)}}}
{N_{{\hat n}}{(a,z)}!}
\end{eqnarray}
where the sum runs over the ensemble $\Pi_{A,Z}$ of different
partitions of $A$ and $Z$ of the  decaying system : $\{{\hat
n}\}=\{N_{{\hat n}}{(1,0)}, N_{{\hat n}}{(1,1)},\dots,N_{{\hat
n}}{(A,Z)}\}$ and $N_{{\hat n}}{(a,z)}$ is the number of
fragments $(a,z)$ in the partition $\{{\hat n}\}$. In this
approximation, the recurrence relation technique
\cite{pgt99,mek93} can be applied providing exact expression for
${\cal Q}_{q}{(A,Z)}$ \cite{gppt00}.

Given the partition function, the mean value of any quantity is
\cite{tsallis} :
\begin{eqnarray}
\label{eq5} <{\cal O}>_q=\sum_{\vec p}{\cal O}_{\vec p}{\hat
p}_{\vec p}^q \ .
\end{eqnarray}
In order to ensure the proper normalization of $q-$averages
(\ref{eq5}), it is better to work with the generalized averages
\cite{tsallis98} :
\begin{eqnarray}
\label{eq10} {\ll{\cal O}\gg}_q=<{\cal O}>_q/<1>_q~.
\end{eqnarray}
These normalized mean values exhibit all convenient properties of
the original mean values. Moreover, the TGSM can be reformulated
in terms of ordinary linear mean values calculated for the
renormalized entropic index : $q^*=1+(q-1)/q$. In particular, the
total average energy and pressure of the system become :
\begin{eqnarray}
\label{eq11}
{\cal E}_{q} &=&
\sum_{a,z}<N{(a,z)}>_{q^* A Z} \  <\varepsilon(a,z)>_{q^*}   \\
P_{q} &=&
\sum_{a,z}<N{(a,z)}>_{q^* A Z} \ <p(a,z)>_{q^*}
\label{eq11a}
\end{eqnarray}
where  $<\varepsilon(a,z)>_{q}$ and $<p(a,z)>_{q}$ are given by :
\begin{eqnarray}
\label{eq7}
<{\varepsilon(a,z)}>_q &=& -\frac{\partial}{\partial \beta}\left(
\frac{1-[\omega_q(a,z)]^{-q_1}}{q_1}\right)    \\
<p(a,z)>_q &=&\frac {1}{\beta } \frac{\partial}{\partial V_f}\left(
\frac{1-[\omega_q(a,z)]^{-q_1}}{q_1}\right)
\label{eq7a}
\end{eqnarray}
 and the average multiplicity of $(a,z)$-fragments in the fragmentation of
system $(A,Z)$ is :
\begin{eqnarray}
\label{eq12}
<N(a,z)>_{q A Z}=\omega_{q}{(a,z)}
\frac{{\cal Q}_{q}{(A-a,Z-z)}}{{\cal Q}_{q}{(A,Z)}}
\end{eqnarray}

The heat capacity at a constant volume ($=\partial {\cal
E}_q/\partial T\mid_{V_f}$) is:
\begin{eqnarray}
\label{eq13}
C_V&=&{\beta}^2\{
\sum_{a,z}\sum_{a^{'},z^{'}}
<\Delta(az;a^{'}z^{'})>_{q^*} \ <\varepsilon(a,z)>_{q^*}\times  \nonumber
\\  &\times& <\varepsilon(a^{'},z^{'})>_{q^*}
+ \sum_{a,z}<N{(a,z)}>_{q^{*} A Z}\times  \nonumber \\  &\times&
[<\varepsilon^2(a,z)>_{q^*}-<\varepsilon(a,z)>_{q^*}^2]\}
\end{eqnarray}
where :
\begin{eqnarray}
&&<\Delta(az;a^{'}z^{'})>_{q}\equiv <N{(a,z)}N{(a^{'},z^{'})}>_{q A
Z}-\nonumber
\\&& - <N{(a,z)}>_{q A Z} \ <N{(a^{'},z^{'})}>_{q A Z}
\end{eqnarray}
and :
\begin{eqnarray}
\label{eq14}
&&<N{(a,z)}N{(a^{'},z^{'})}>_{q A Z}=\nonumber \\
&&\omega_q{(a,z)}\omega_q{(a^{'},z^{'})}
\frac{{\cal Q}_q{(A-a-a^{'},Z-z-z^{'})}}{{\cal Q}_q{(A,Z)}}+\nonumber \\
&+&\delta_{aa^{'}}\delta_{zz^{'}}\omega_q{(a,z)}\frac{{\cal
Q}_q{(A-a,Z-z)}}{{\cal Q}_q{(A,Z)}} ~ \ .
\end{eqnarray}
The heat capacity at a constant pressure $C_P$
 ($=\partial ({\cal E}_q+{\cal P}_qV_f)/\partial T\mid_{{\cal P}_q}$)
can be calculated using the relation :
$C_P-C_V=T V_f \ \kappa_T \ (\partial {\cal P}_q/\partial T\mid_{V_f})^2~ \ ,$
where $\kappa_T$ stands for the isothermal compressibility
($=-(1/V_f)\partial V_f/\partial {\cal P}_q\mid_T$) :
\begin{eqnarray}
\frac{1}{\kappa_T}&=&-\beta V_f\left[
\sum_{a,z}\sum_{a^{'},z^{'}}
<\Delta(az;a^{'}z^{'})>_{q^*}\times \right. \nonumber \\ &\times&
<p(a,z)>_{q^*} \
 <p(a^{'},z^{'})>_{q^*} +       \\
&+& \left. \sum_{a,z}<N{(a,z)}>_{q^{*} A Z}
\frac{1}{\beta }
\left( \frac{\partial <p(a,z)>_{q^*}}{\partial V_f} {\mid}_T \right)
\right]                        \nonumber
\end{eqnarray}
and :
\begin{eqnarray}
&&\frac{\partial {\cal P}_q}{\partial T}{\mid}_{V_f} = \beta^2 \left[
\sum_{a,z}\sum_{a^{'},z^{'}}
<\Delta(az;a^{'}z^{'})>_{q^*} \times \right. \nonumber \\
&\times&  <p(a,z)>_{q^*} \
 <\varepsilon(a^{'},z^{'})>_{q^*}+  \\
&+& \left. \sum_{a,z}<N{(a,z)}>_{q^{*} A Z}
\frac{1}{\beta^2 }\left( \frac {\partial <p(a,z)>_{q^*}}{\partial T}
{\mid}_{V_f} \right) \right] ~  \nonumber
\end{eqnarray}
One should stress that all these thermodynamical quantities are
calculated {\it exactly}, without using the Monte Carlo technique.

The upper part of Fig. 1 shows the temperature dependence of the
pressure for various entropic indices $q$ in systems with
$A_0=100, 200$ and $300$ nucleons and $Z_0=0.4A_0$ protons. In
the bottom part, the temperature dependence of the inverse
thermal compressibility  $1/\kappa_T$ is shown. Zero of $1/
\kappa_T$ ($\partial{{\cal P}_q}/\partial{V_f}\mid_T=0$)
corresponds to the pole of $C_P$ and defines the boundary of the
two phase coexistence region. For $q>1$, there exists a region of
temperatures where $1/\kappa_T$ is negative and, hence, $C_P$
becomes negative between the poles. In the BGSM limit,
$1/\kappa_T$ has  zeros  for  $A_0=100, 200$, whereas in heavier
systems these zeros appear only for $q>1$. An essential part of
the pressure and, hence, of $1/\kappa_T$ is the Coulomb term. The
inverse compressibility $1/\kappa_T$ never vanishes when the
Coulomb term is neglected \cite{pgt99}. Since the Coulomb
contribution to the pressure and the inverse thermal
compressibility decreases in the Wigner-Seitz approximation
roughly as $A_0^{-1/3}$, this particular signature may not be
seen in heavy systems in the BGSM limit. Existing data do not
allow yet to pin down the $A_0-$dependence of the 'criticality'
signatures. Nevertheless, Fig. 1 demonstrates how fragile is the
Boltzmann-Gibbs equilibrium 'critical' behavior. Small increase
of $q$ above  the BGSM limit leads to an upwards shift  of the
critical temperature $T_c$ which, for the same value of $q$, is
higher in heavier systems. All these important changes take place
in a narrow range of temperatures around $T_c$, beyond which the
fragmenting system closely follows the BGSM limit.
\vspace*{-15mm}

\begin{figure}[h]
\epsfig{figure=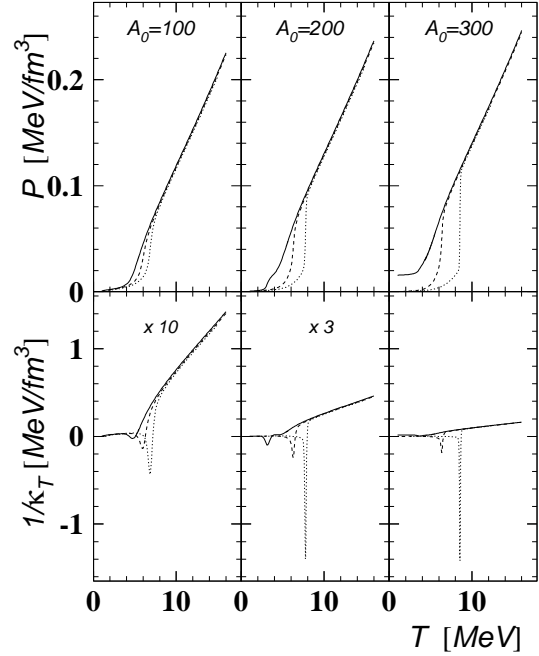,height=11cm} \caption{The dependence
of the pressure (the upper part) and the inverse isothermal
compressibility (the lower part) on the temperature $T$ is plotted
for system of different sizes  and different entropic indices $q$
: 1.0 (the solid line), 1.0005 (the dashed line), 1.001 (the
dotted line). The freeze-out volume $V_f$ corresponds to
$\rho_f\equiv A_0/V_f=\rho_0/4$. The calculated values of $1/k_T$
are multiplied by factors 10 and 3 for $A_0=$100 and 200,
respectively.}
\label{fig1}
\end{figure}

Fig. 2 present the heat capacities $C_V$ and $C_P$ as a function
of the average excitation energy : $E^*={\cal E}_q(T,V_f) - {\cal
E}_q(T=0,V_0)$, where $V_f$ is the freeze-out volume,
$V_0=A_0/\rho_0$ and $\rho_0$ is the equilibrium density at $T=0$.
$C_V$ is a smooth positive function of the excitation energy for
all values of $q$. The peak of $C_V(E^*)$, whose position is
associated with the critical temperature $T_c$, becomes more
pronounced for higher $q$. Fig. 3 compares the heat capacity
$C_P$ vs $E^*/A_0$ for systems of different sizes : $A_0=200$ and
$A_0=300$. In the BGSM limit, the negative branch of $C_P$ is
seen only for $A_0 \lsim 200$. With increasing $A_0$, its
position moves towards lower excitation energies. It should be
noted that the critical density $\rho_c$ in the nonextensive
fragmentation model \cite{gppt00} is relatively high. The
'global' critical point $(V_c,T_c,P_c)$ for $A_0=100$ corresponds
to $\rho_c/\rho_0=0.547, 0.783$ and 0.925 for $q=1, 1.0005$ and
$q=1.001$, respectively. For $A_0=200$, the global critical point
exists only for $q=1$ whose value of $\rho_c/\rho_0=0.904$ is
close to that obtained in statistical multifragmentation model
using the recurrence relation technique~\cite{EH00}. For $q>1$,
the system is always found inside of the two phase coexistence
region for all $\rho < \rho_0$. This is consistent with the
picture of nonextensivity driven first order phase transition.

\begin{figure}[h]
\vspace*{-15mm} \epsfig{figure=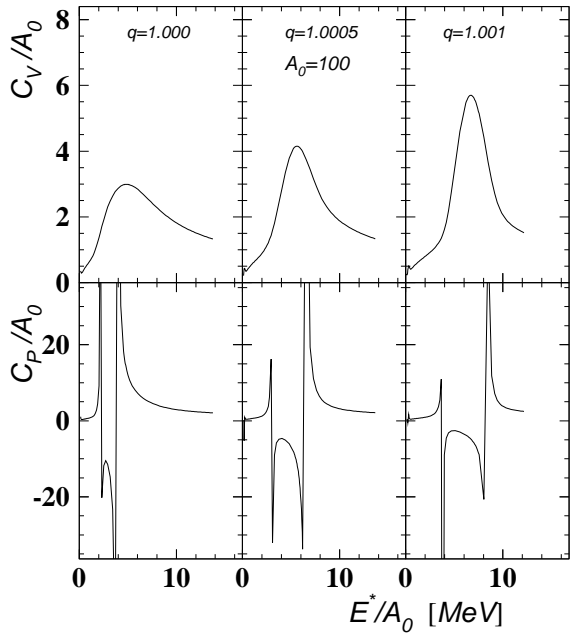,height=10.5cm}
\caption{The specific heat at a constant volume $C_V$ (the upper
part) and at a constant pressure $C_P$ (the lower part) are
plotted vs the excitation energy per nucleon for various entropic
indices $q$ in the system with $A_0=100$ and $Z_0=40$. }
\label{fig2}
\end{figure}

The description of nuclear matter in terms of the Van der Waals
fluid \cite{EH00} (see also \cite{JMZ83}) yields much lower
critical densities ($\rho_c \approx 0.3\rho_0$). In this model,
the boundary of the coexistence region has a bell-like shape and
the line $V=V_f$ crosses it in a single point. Consequently, the
negative branch of heat capacity is not seen. In the nonextensive
fragmentation model \cite{gppt00} , the boundary of the
coexistence region is skew with the top tilted towards smaller
$V$ what allows for two crossings with the line $V=V_f$ and leads
to the negative branch of $C_P$.

In conclusion, the phase transition in the statistical nuclear
multifragmentation models tends to disappear in heavy systems due
to the weakening of the Coulomb contribution. This effect can be
compensated by the nonextensive features of entropy due to either
long-range correlations/memory effects or the fractality of the
liquid-gas interphase, which both tend to strengthen signatures
of the first order phase transition. The application of
nonextensive canonical statistical  fragmentation model
\cite{gppt00} for the understanding of experimental 'caloric'
curve data \cite{exp1} and the negative heat capacity data
\cite{expCV} in the 'critical' region, consistently indicates
deviation from the BGSM picture of the phase transition and
$q\gsim 1.0005$. This tiny variation of $q$, which cannot be
detected either in the particle/fragment kinetic energy
distributions or in the angular distributions, have strong
measurable effects on the event-by-event energy fluctuations of
particles/fragments in the region of phase coexistence. Hence,
the mass-dependence of the criticality signatures is determined
by a subtle competition between Coulomb repulsive interactions
and nonextensive features of the entropy.
\begin{figure}[h]
\vspace*{-15mm} \epsfig{figure=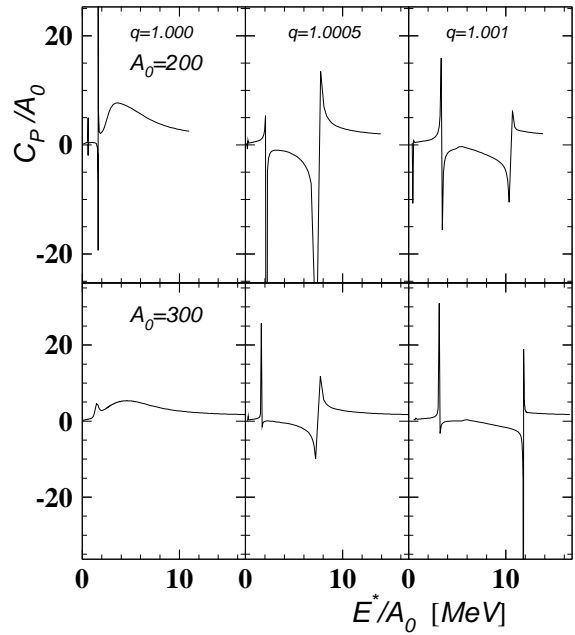,height=10.5cm}
\caption{The specific heat at a constant pressure $C_P$ is
plotted vs the excitation energy per nucleon for various entropic
indices $q$ in the systems with $A_0=200$, $Z_0=80$ (the upper
part) and $A_0=300$, $Z_0=120$ (the lower part).} \label{fig3}
\end{figure}

For $q>1$, the negative branch of $C_P$ is seen both in light and
heavy systems. The range of excitation energies corresponding to
$C_P<0$  increases with increasing $A_0$. However, in heavy
systems the negative branch of $C_P$ appears {\it uniquely} for
$q>1$. Both extension and localization of the negative branch of
$C_P$ in quasi-peripheral Au+Au collisions at 35A.MeV
\cite{expCV}, closely resemble results of nonextensive
fragmentation model for $q\simeq 1.0005$ and $A_0=200$ (see Fig.
3). This suggests that the observed effect is caused primarily by
the nonextensive features of the entropy. The position of
singularity of $C_P$ at higher excitation energies increases
sensitively both with the entropic index $q$ and with the source
size. Correct description of its experimental value \cite{expCV}
by the nonextensive fragmentation model assuming maximal possible
size of quasi-projectile source in this experiment ($A_0\lsim
200$) \cite{expCV}, means that we have determined lower limit for
$q$.

There are many sources of nonextensivity in mesoscopic systems. Some of them
, {\it e.g.} the formation of liquid-gas (fractal) interphase \cite{Gross},
have been pointed out in the microcanonical studies \cite{Gross,many1}.
Most of these effects are non-generic and, moreover, they are
hard to quantify. This is a principal obstacle in the meaningful
characterization of nuclear multifragmentation data in the 'critical region'
in the idealized picture of BGSM.
On the other hand, the same value of entropic index
($q\sim 1.0005$) seems to be consistent with both the
caloric curve \cite{exp1} and the negative heat capacity \cite{expCV}
data, in spite of
completely different kinematical conditions in these measurements.
Moreover, the excitation energy of higher
singularity of $C_P$ seems to be the same both in quasi-peripheral
Au+Au collisions at 35A.MeV \cite{expCV}
and in central Xe+Sn collisions at 32A.MeV \cite{indra}
and agrees with $q\simeq 1.0005$.
This surprising universality remains a puzzle at present.

\vspace{0.2cm}
\noindent
{\bf Acknowledgements}\\
 The work was supported by
the IN2P3-JINR agreement No 00-49.

\end{document}